\documentclass[12pt,a4paper]{article}

\usepackage{amssymb}
\usepackage{epsfig}
\usepackage[figuresright]{rotating}

\begin{document}
\noindent
DESY 03-177, SFB/CPP-03-52\hfill{\tt hep-lat/0311023}\\
November 2003
\vspace{50pt}

\begin{center}{\Large\bf The pion mass in finite volume}
\end{center}
\vspace{20pt}

\begin{center}
{\large\bf Gilberto Colangelo$\,{}^{a}$}
{\large and}
{\large\bf Stephan D\"urr$\,{}^{b}$}
\\
\vspace{10pt}
${}^a${\small\it Institute for Theoretical Physics, University of Bern,
                 3012 Bern, Switzerland}\\
${}^b${\small\it DESY Zeuthen, 15738 Zeuthen, Germany}\\
\end{center}
\vspace{20pt}

\begin{abstract}
\noindent
We determine the relative pion mass shift $M_\pi(L)/M_\pi-1$ due to the finite
spatial extent $L$ of the box by means of two-flavor chiral perturbation
theory and the one-particle L\"uscher formula. We use as input the
expression for the infinite volume $\pi \pi$ forward scattering amplitude
up to next-to-next-to-leading order and can therefore control the
convergence of the chiral series. A comparison to the full leading order
chiral expression for the pion mass in finite volume allows us to check the
size of subleading terms in the large-$L$ expansion.
\end{abstract}

\clearpage


\newcommand{\mpi}{M_{\pi}}
\newcommand{\fpi}{F_{\pi}}
\newcommand{\Mpi}{M_{\pi}}
\newcommand{\Fpi}{F_{\pi}}

\newcommand{\ovr}{\over}
\newcommand{\til}{\tilde}
\newcommand{\pri}{^\prime}
\renewcommand{\dag}{^\dagger}
\newcommand{\<}{\langle}
\renewcommand{\>}{\rangle}
\newcommand{\gaf}{\gamma_5}
\newcommand{\lap}{\triangle}
\newcommand{\trc}{\rm tr}

\newcommand{\al}{\alpha}
\newcommand{\be}{\beta}
\newcommand{\ga}{\gamma}
\newcommand{\de}{\delta}
\newcommand{\ep}{\epsilon}
\newcommand{\ve}{\varepsilon}
\newcommand{\ze}{\zeta}
\newcommand{\et}{\eta}
\renewcommand{\th}{\theta}
\newcommand{\vt}{\vartheta}
\newcommand{\io}{\iota}
\newcommand{\ka}{\kappa}
\newcommand{\la}{\lambda}
\newcommand{\rh}{\rho}
\newcommand{\vr}{\varrho}
\newcommand{\si}{\sigma}
\newcommand{\ta}{\tau}
\newcommand{\ph}{\phi}
\newcommand{\vp}{\varphi}
\newcommand{\ch}{\chi}
\newcommand{\ps}{\psi}
\newcommand{\om}{\omega}

\newcommand{\beq}{\begin{equation}}
\newcommand{\eeq}{\end{equation}}
\newcommand{\bea}{\begin{eqnarray}}
\newcommand{\eea}{\end{eqnarray}}
\newcommand{\bdm}{\begin{displaymath}}
\newcommand{\edm}{\end{displaymath}}

\newcommand{\ltil}{\tilde{\ell}}
\newcommand{\rtil}{\tilde{r}}
\newcommand{\lhat}{\hat{\ell}}
\newcommand{\Ltil}{\tilde{L}}

\newcommand{\mr}{\mathrm}
\newcommand{\mb}{\mathbf}
\newcommand{\Nf}{N_{\!f\,}}
\newcommand{\Nc}{N_{\!c\,}}
\newcommand{\MeV}{\,\mr{MeV}}
\newcommand{\GeV}{\,\mr{GeV}}
\newcommand{\TeV}{\,\mr{TeV}}
\newcommand{\fm}{\,\mr{fm}}
\newcommand{\MSbar}{{\overline{\mr{MS}}}}

\newcommand{\lr}[1]{\ell^{\mr r}_{#1}}
\newcommand{\li}{\ell_i}
\newcommand{\lb}{\bar\ell}
\newcommand{\rt}{\tilde r}
\newcommand{\pa}{\partial}
\newcommand{\nt}{\tilde\nu}
\newcommand{\nn}{\nonumber\\}
\def\fs{\; \; .}
\def\co{\; \; ,}
\def\bb{\bar{b}}
\def\bbar{\bar{b}}
\def\gev{\,\mr{GeV}}

\hyphenation{author another experi-ments scatte-ring rele-vant}


\section{Introduction}


Shifts in particle masses and matrix elements due to the finite extent of
the box are systematic effects in the Monte Carlo treatment of any lattice
field theory.  These shifts become particularly large when the spectrum
contains light particles, and apply to all particles, no matter how heavy,
provided they couple to the light ones. In QCD with light quarks, e.g.\ the
nucleon mass receives such a correction, which disappears only as the product
of the \emph{pion} mass times the length of the box ($\Mpi L$) gets large.

Fortunately, as long as pions are light, chiral symmetry imposes strong
constraints on the way observables deviate from their infinite-volume limit:
chiral perturbation theory (CHPT) allows one to perform a systematic
expansion around this limit and hence to control the finite-volume effects
analytically \cite{GL84,GaLeFSE1,Gasser:1987zq,Leutwyler:1987ak,betterways}.
In some cases one can directly translate the results of finite volume
simulations into information about infinite-volume quantities without any 
need to extrapolate in the box size $L$~\cite{eps_exp_use}.
CHPT is an expansion in the pion mass and particle momenta which have
to be small in comparison to the chiral symmetry breaking scale, usually
identified with $4 \pi F_\pi$. The conditions of applicability read
\beq
{p\over4\pi\fpi}\ll1\;,\quad{\mpi\over4\pi\fpi}\ll1
\label{xpt_conditions}
\;.
\eeq
Particles inside a box of spatial length $L$ with periodic boundary conditions
and comparatively large or infinite time extent
may only have discrete values of their spatial momenta,
$p_k=2\pi n_k/L$ with $n_k\in\mathbf{Z}$. In this case the first condition
in (\ref{xpt_conditions}) becomes a bound on the box size:
\beq
L\gg{1\over2F_\pi}\sim1\fm
\;.
\label{xpt_cond_mod}
\eeq
Note that these conditions have no say whether $L$ is large compared to the
Compton wavelength of the pion or not.
Both options are acceptable \cite{betterways}, but they imply
different ways to organize the chiral series:
\bea
\mpi L \gg 1 \quad \leftrightarrow \quad \mbox{``$p$-expansion''}\,\,
\label{xpt_mom}
\\
\mpi L \ll 1 \quad \leftrightarrow \quad \mbox{``$\epsilon$-expansion''}\;.
\label{xpt_eps}
\eea
We shall here restrict ourselves to the former case but investigate what
the conditions (\ref{xpt_conditions}b, \ref{xpt_cond_mod}) mean
quantitatively, by considering more than the leading order in the chiral
expansion. The underlying assumption is that CHPT itself will tell us when
the conditions of applicability are not respected any more, via a bad
convergence behavior.

\begin{figure}[b]
\hfill
\unitlength 0.5cm
\begin{picture}(14,8)(0,0)
\linethickness{0.2mm}
\put(0,0){\line(0,1){8}}
\put(5,0){\line(0,1){8}}
\put(0.5,7.8){\line(1,0){4}}
\put(0.25,7.65){$\blacktriangleleft$}
\put(4.25,7.65){$\blacktriangleright$}
\put(2.25,7.2){$L$}
\multiput(2.04,6.6)(0.1,0){10}{\circle*{0.5}} 
\multiput(2.04,0.5)(0.1,0){10}{\circle*{0.5}} 
\linethickness{0.8mm}
\multiput(2.5,0.75)(0,0.85){7}{\line(0,1){0.5}}
\multiput(0.55,3.55)(0.85,0){5}{\line(1,0){0.5}}
\put(0,3.55){\line(1,0){0.25}}
\put(4.75,3.55){\line(1,0){0.25}}
\linethickness{0.5mm}
\put(12.05,4){\circle{2}}
\linethickness{1.5mm}
\put(10.5,3.5){\line(1,1){1}}
\put(10.5,4.5){\line(1,-1){1}}
\linethickness{0.2mm}
\put(13.1,1){\line(0,1){6}}
\end{picture}
\hfill
\vspace{-2pt}
\caption{The mass shift due to quantized momenta in the self-energy corrections
amounts to a finite-size effect from pion exchange ``around the world'' (left),
depicted by a ``thermal insertion'' (cross) in diagrammatic language (right).}
\label{fig:around}
\end{figure}
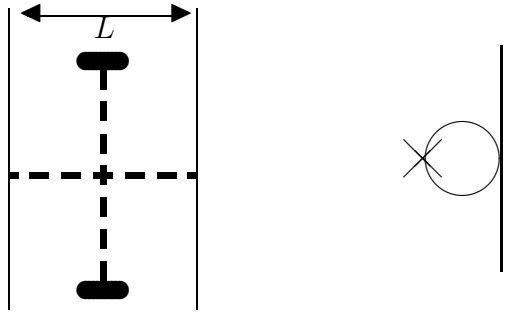

In this paper we study the pion mass, $\mpi(L)$, defined as eigenvalue of
the QCD Hamiltonian in a $L\!\times\!L\!\times\!L$ box (with periodic
boundary conditions), as it is extracted on an Euclidean lattice for
sufficiently large time $T$. The finite volume shift, $\Mpi(L)-\Mpi$, is
--~in coordinate space view~-- due to the possibility of the pion wrapping
``around the world'' or --~in momentum space view~-- due to the discrete
momenta, as depicted in Fig.\,\ref{fig:around}.  The ultraviolet properties
of the theory are untouched -- with toroidal boundary conditions no new
counterterms are introduced and the finite size effect is automatically
finite. Our goal is to study the shift $\Mpi(L)-\Mpi$ (with $\Mpi$ we always
denote the infinite-volume value) as a function of $\Mpi$ and $L$.  A
preliminary account has been given in \cite{Colangelo:2002hy}. Here, we
include one more order in the chiral expansion (at large $L$) and carefully
analyze the uncertainty due to the errors of the QCD low-energy constants
involved. In addition, we compare to the full one-loop result by Gasser and
Leutwyler (which is leading in the chiral expansion, but includes the
large-$L$ suppressed terms).


\section{The pion mass in finite volume}


\subsection{Finite volume calculations in CHPT}

In a series of remarkable papers
\cite{GaLeFSE1,Gasser:1987zq,Leutwyler:1987ak}, Gasser and 
Leutwyler have shown that chiral symmetry does severely constrain physical
observables at low energy even if the system is confined to a finite
box. The only condition is that the box length $L$ must be several fermi,
cf.\ (\ref{xpt_cond_mod}).

Using the method of the effective chiral Lagrangian, one can establish low
energy theorems for physical quantities of interest, which have the generic
form
\bea
Q&=&Q_0 \left[1+ \xi q_1 + \xi^2 q_2 +O(\xi^3) \right]
\label{QL}
\\
\xi&\equiv&{\Mpi^2\ovr(4\pi\Fpi)^2} \co
\label{defxi}
\eea
and the coefficients $q_i$ are quantities of order one in the chiral expansion.
They depend on the low--energy constants (LEC), on ratios of momenta and quark
masses, and, in finite volume, on
\beq
\lambda\equiv\Mpi L
\fs
\label{defla}
\eeq

\begin{table}[b]
\centering
\begin{tabular}{|l|rrrrrrrrrrrrrrrrrrrr|}
\hline
$n$&1&2&3&4&5&6&7&8&9&10&11&12&13&14&15&16&17&18&19&20\\
$m(n)$&6&12&8&6&24&24&0&12&30&24&24&8&24&48&0&6&48&36&24&24\\
\hline
\end{tabular}
\caption{The multiplicities $m(n)$ in (\ref{g1til2}) for $n\leq20$.}
\label{tab:mN}
\end{table}

For the pion mass and decay constant, the coefficient $q_1$ in the
$p$-expansion (\ref{xpt_mom}) has been explicitly evaluated in
Ref.\,\cite{GaLeFSE1},
\bea
\Mpi(L)&=&\Mpi\left[1+{1\ovr2\Nf}\xi\,\tilde g_1(\lambda)+O(\xi^2)\right]
\label{mpi}
\\
\Fpi(L)&=&\;\Fpi\left[1-\;{\Nf\over2}\,\xi\,\tilde g_1(\lambda)+O(\xi^2)\right]
\label{fpi}
\eea
with
\beq
\tilde g_1(\lambda) = \sum\nolimits' \int_0^\infty dx \;
e^{-{ 1 \over x}-{x\over 4}(n_1^2+n_2^2+n_3^2)\la^2} \co
\label{gtil}
\eeq
where the sum runs over all integer values of $n_{1,2,3}$, excluding
the term with $(n_1,n_2,n_3)=(0,0,0)$ as indicated by the prime over the
summation symbol. For a given value of $n:=n_1^2+n_2^2+n_3^2$ the
integral can be performed analytically, and we can rewrite $\tilde g_1$ as
\beq
\tilde g_1(\la) = \sum_{n=1}^\infty
{4 m(n)\ovr\sqrt{n}\,\la}\; K_1\left(\sqrt{n}\,\lambda\right) \co
\label{g1til2}
\eeq
where $K_1$ is a Bessel function of the second kind and the multiplicity $m(n)$
indicates how many times the value $n$ is generated in the sum in
Eq.\,(\ref{gtil}). The values of $m(n)$ for $n\leq20$ are given in
Table~\ref{tab:mN}. 
For large argument the Bessel function $K_1$ drops exponentially:
$K_1(x) \simeq e^{-x}/\sqrt{x}$. Since $\la\!\gg\!1$ by assumption, the sum
(\ref{g1til2}) converges very rapidly, and it is easy to check how many
terms are needed to get a good approximation for the complete sum. The
convergence of the chiral expansion, on the other hand, is more difficult to
test: the coefficient $q_2$ has not yet been calculated neither for $\Mpi$ nor
for $\Fpi$. This would require a full two--loop calculation in CHPT in finite
volume. We are going to argue, however, that there is a fast and reliable way
to check the convergence of the chiral expansion -- a way that is based on
a formula due to L\"uscher \cite{Luscher:1985dn} which we will
discuss in detail. Before doing so, it is useful to rearrange 
Eq.\,(\ref{mpi}, \ref{fpi}) by making explicit the
expansion of each coefficient $q_m$ in a series of rapidly decreasing
exponentials (in $\lambda$)
\bea
\Mpi(L) &=&\Mpi
\Big[1+ \left(\xi q^M_{11}+\xi^2 q^M_{21}+O(\xi^3)\right) K_1(\la) \nn
& & \qquad\;\;\,
+\left(\xi q^M_{12}+\xi^2 q^M_{22}+O(\xi^3)\right) K_1(\sqrt{2}\la)
+\;\ldots \Big] \nn
&=&\Mpi
\left[1+ \sum_{m=1}^\infty \sum_{n=1}^\infty
\xi^m q^M_{mn} \,K_1(\sqrt{n}\,\la) \right]
\label{MpiL}
\\
\Fpi(L)&=&\,\Fpi\,
\left[1+ \sum_{m=1}^\infty \sum_{n=1}^\infty
\xi^m q^F_{mn} \,K_1(\sqrt{n}\,\la) \right] \co
\label{FpiL}
\eea
where at leading order in the chiral expansion the coefficients $q_{1n}$ follow
from Table 1,
\beq
q^M_{1n}=+{2m(n)\ovr\Nf\sqrt{n}\,\la}\co \qquad \qquad
q^F_{1n}=-{2\Nf m(n)\ovr\sqrt{n}\,\la} \fs
\label{q1MF}
\eeq
Notice that the Bessel functions that we have factored out in
Eqs.\,(\ref{MpiL}, \ref{FpiL}) need not necessarily appear in this form also
at higher orders in the chiral expansion -- the coefficients $q^{M,F}_{mn}$
are presumably complicated functions of $\lambda$, in general. We do
expect, however, that also at higher orders the result may be expressed as
a series of exponentials, whose leading asymptotic
behavior should be captured by the Bessel functions in
(\ref{MpiL},\ref{FpiL}). 


\subsection{L\"uscher's formula for $\Mpi(L)$}

\begin{figure}[tb]
\hfill
\unitlength 0.6cm
\begin{picture}(12,8)(0,0)
\linethickness{0.2mm}
\put(0,3){\line(1,0){6.5}}
\put(7.2,3){\line(1,0){4.8}}
\put(6,0){\line(0,1){6.5}}
\put(1,6){\oval(1.5,1.5)}
\put(0.85,5.85){$\nu$}
\linethickness{1.6mm}
\put(00,3){\line(1,0){2}}
\put(10,3){\line(1,0){2}}
\put(1.0,2){\large $-\Mpi$}
\put(9.5,2){\large $ \Mpi$}
\linethickness{0.6mm}
\put(6,0.5){\line(0,1){5}}
\put(5.75,5.5){$\blacktriangle$}
\put(6.5,1.4){\begin{turn}{90}{\normalsize\bf Integration}\end{turn}}
\end{picture}
\hfill
\vspace{-10pt}
\caption{Integration contour in the complex $\nu$ plane where $\nu$ is the
crossing variable in the (Minkowski space) forward scattering amplitude.
The cut is generated only at NLO in the chiral expansion.}
\label{fig:integration}
\end{figure}
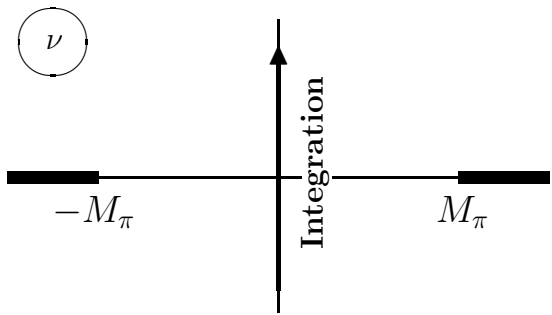

L\"uscher considered the problem of evaluating the finite size effects on a
particle mass from a different point of view.  Via graph--theoretic
arguments he proved an elegant relation between the (Euclidean) finite
volume mass shift and the (Minkowski space) $\pi \pi$ forward scattering
amplitude $F(\nu)$ in infinite volume \cite{Luscher:1985dn}:
\begin{equation}
\Mpi(L)-\Mpi =-{3\over16\pi^2\Mpi L}\;\int_{-\infty}^\infty\!dy\; 
F({\rm i}y)\,e^{-\sqrt{\Mpi^2+y^2}\,L}+O(e^{-\overline{M}L}) \fs
\label{luscher}
\end{equation}
The integration runs along the imaginary axis, where the amplitude is far
away from its cuts (cf.\ Fig.\,\ref{fig:integration}): only the real part
of $F({\rm i}y)$ contributes to the integral. The one-particle
L\"uscher formula (\ref{luscher}) is an asymptotic relation for large $L$;
it is proven to all orders in perturbation theory for an arbitrary massive
QFT, and the subleading piece is tamed by the bound
$\overline{M}\!\geq\!\sqrt{3/2}\:\Mpi$ \cite{Luscher:1985dn}. An extra term
originating from the 3-particle vertex in the original formula
\cite{Luscher:1985dn} does not appear in our case, due to the odd intrinsic
parity of the pion and parity conservation in QCD. Since
eq.\,(\ref{luscher}) builds on the unitarity of the theory, it holds only
in full QCD. A variant for the quenched approximation is not known, if at
all possible.

Neither the formula, nor its derivation make use of the chiral expansion
(indeed the formula was derived before CHPT was applied to finite volume
effects). However, for practical applications we need to insert an
explicit, analytic representation for the $\pi \pi$ forward scattering
amplitude, and for this we shall rely on CHPT. A comparison to the
expression (\ref{MpiL}) shows that by inserting the chiral expansion for
$F$ in (\ref{luscher}) and evaluating the integral one can obtain the
coefficients $q^M_{m1}$.  If we write (extracting a factor $16 \pi^2$ for
later convenience)
\beq
F({\rm i}y)=16\pi^2 \left[\xi F_2+\xi^2 F_4+\xi^3 F_6+O(\xi^4)\right] \co
\label{forw}
\eeq
the integration of the term $F_{2m}$ yields the coefficient $q^M_{m1}$.
Since for 2 degenerate dynamical flavors the $\pi \pi$ scattering amplitude
is known to next-to-next-to-leading order \cite{BCEGS} in the chiral
expansion, i.e.\ up to $F_6$, we can evaluate the coefficients $q^M_{m1}$,
$m\!=\!1,2,3$ for $\Nf\!=\!2$. All this without having to do a single new
loop calculation in CHPT in finite volume.

We stress that the L\"uscher formula is designed to yield the {\em leading
term\/} in the large $L$ expansion: whenever the relative suppression factor
\beq
{\mbox{subleading}\ovr\mbox{leading}}=O\Big(e^{-(\overline{M}-\Mpi)L)}\Big)
\eeq
is not small, subleading effects may be of practical relevance. In order to
estimate subleading effects in the large $L$ expansion we will go back to the
expansion of $\til g_1(\lambda)$.

The rest of this article is devoted to such a combined use of L\"uscher's
formula (\ref{luscher}) and the chiral expansion (\ref{forw}) for
two-flavor QCD. It is useful to keep in mind that the chiral formulae
contain low-energy constants that have been pinned down in real Minkowski
space experiments; some stem indeed from $\pi\pi$ scattering data, which
determine the amplitude in the physical region, e.g.\ for forward
kinematics, on the cuts in Fig.\,\ref{fig:integration}. The very fact that
the chiral expression for $F(\nu)$ holds in the whole complex $\nu$ plane
is a key ingredient of this work -- we rely on the good analyticity
properties of the chiral amplitudes.


\subsection{Combining CHPT and the L\"uscher formula}

The forward scattering amplitude which is needed in L\"uscher's formula can
be expressed in terms of the isospin invariant amplitude $A(s,t,u)$ as follows 
\bea
F(\nu) &=& T(2\mpi(\mpi+\nu),0,2\mpi(\mpi-\nu)) \nn
T(s,t,u)&=&A(s,t,u)+3A(t,s,u)+A(u,s,t) \co
\eea
where the only independent kinematical variable $\nu$ reads
\begin{equation}
\;\;\nu \equiv {p_a \cdot p_b \over \Mpi}={s\over2\Mpi}-\Mpi \co
\label{cross}
\end{equation}
in terms of the momenta $p_{a,b}$ of the two initial pions.
Since $A(s,t,u)$ is symmetric under $t,u$ exchange, the forward scattering
amplitude is an even function of $\nu$:
$F(\nu)\!=\!F(-\nu)$.

\begin{figure}[t!]
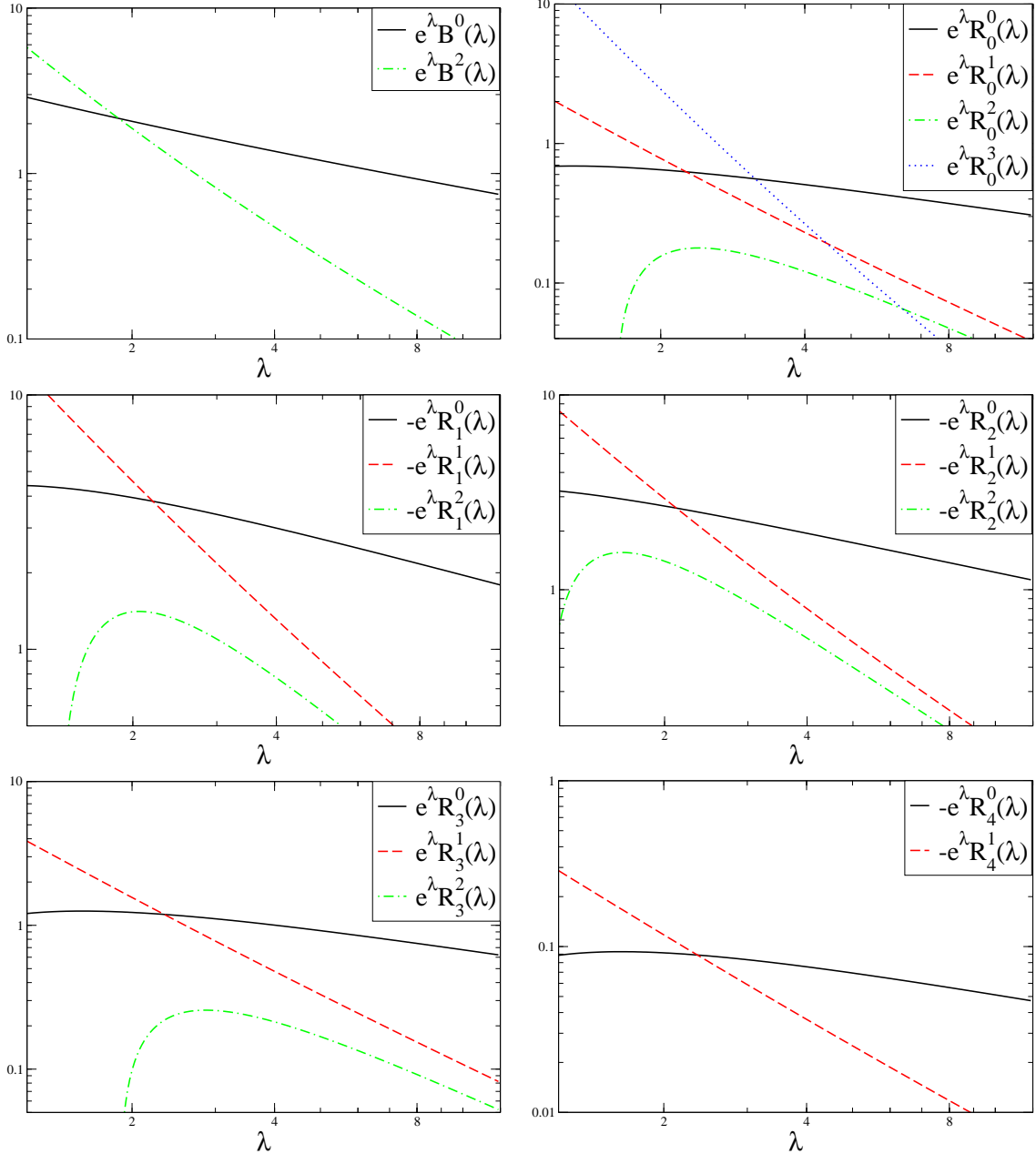

\begin{center}
\begin{tabular}{rr}
\epsfig{file=Bi,height=5.5cm}&\epsfig{file=R0,height=5.56cm}\\
\epsfig{file=R1,height=5.5cm}&\epsfig{file=R2,height=5.5cm}\\
\epsfig{file=R3,height=5.5cm}&\epsfig{file=R4,height=5.5cm}
\end{tabular}
\vspace{-6mm}
\end{center}
\caption{The integrals $B^{2k}$ and $R_i^k$ which are needed for the evaluation
of the finite volume corrections to NNLO. The $B^{2k}$ are known analytically,
the $R_i^k$ have been determined numerically.}
\label{fig:BR}
\end{figure}

We find it convenient to use a dimensionless integration variable and introduce
\beq
\til y=y/\mpi \co \qquad \til\nu = \nu/\mpi \fs
\eeq
In these variables and with $N\!=\!16\pi^2$ the chiral expansion of the
forward scattering amplitude is
\beq
F(\nu)=N\left[ \xi F_2(\til\nu) + \xi^2 F_4(\til\nu) + \xi^3 F_6(\til\nu) +
O(\xi^4) \right] \fs
\eeq
If we combine L\"uscher's formula with the chiral expansion for $F$ we obtain
a simple expression:
\bea
{\mpi(L)-\mpi\ovr\mpi} &\equiv&R_M(\mpi,L) 
  \co \nn
R_M(\mpi,L) &=& -{3 \over \la}
\int_{-\infty}^{\infty}d\til y\;e^{-\sqrt{1+\til y^2}\,\la}
\left[\xi F_2({\rm i}\til y)+ \xi^2
  F_4({\rm i}\til y) + \xi^3 F_6({\rm i}\til y) + O(\xi^4) \right] + \ldots \nn
&=& -{3  \over \la } \left[\xi I_2(\la ) + \xi^2
  I_4( \la ) + \xi^3 I_6(\la) + O(\xi^4) \right] +\ldots \co
\label{RMser}
\eea
where the ellipsis indicate terms of order $e^{-\bar\Mpi L}$ and
\beq
I_{2m}(\la)=\int_{-\infty}^{\infty}d\til y\;
e^{-\sqrt{1+\til y^2}\,\la}\,F_{2m}({\rm i}\til y)
\fs
\eeq
Using the expression for $A$ in \cite{BCEGS} (which is specific to $\Nf=2$)
and splitting its $b_i$ coefficients
\beq
\begin{array}{lcll}
b_i&\equiv&\bb_i/N&\mr{for}\quad i=1,\ldots,4
\\
b_i&\equiv&\bb_i/N^2&\mr{for}\quad i=5,6
\\
\bb_i&\equiv&\bb_i^0+\xi\bb_i^1&\mr{for}\quad i=1,\ldots,4 \co
\end{array}
\label{bisplit}
\eeq
the calculation of the coefficients $I_{2m}$ in (\ref{RMser}) can be made in
large parts analytically:
\bea
I_2(\la)&=&-B^0(\la)
\nn
I_4(\la)&=&B^0(\la)\Big(5\bb_1^0+4\bb_2^0+8\bb_3^0+8\bb_4^0\Big)
\nn
&+&B^2(\la)\Big(-8\bb_3^0-56\bb_4^0\Big)
\nn
&+&{13\over3}R_0^0(\la)-{16\over3}R_0^1(\la)-{40\over3}R_0^2(\la)
\nn
I_6(\la)&=&B^0(\la)\Big(5\bb_1^1+4\bb_2^1+8\bb_3^1+8\bb_4^1+16\bb_5+16\bb_6\Big)
\nn
&+&B^2(\la)\Big(-8\bb_3^1-56\bb_4^1-48\bb_5+16\bb_6\Big)
\nn
&+&R_0^0(\la)\Big(50+10\bb_1^0+{56\over3}\bb_2^0+{104\over3}\bb_3^0+
                  {56\over3}\bb_4^0\Big)\nn
&+&R_0^1(\la)\Big(-{1402\over27}-{32\over3}\bb_2^0-{128\over3}\bb_3^0-
                  {32\over3}\bb_4^0\Big)\nn
&+&R_0^2(\la)\Big(-{1756\over27}-{80\over3}\bb_2^0-{392\over3}\bb_3^0+
                  {136\over3}\bb_4^0\Big)\nn
&+&R_0^3(\la)\Big(-{116\over27}+16\bb_3^0-48\bb_4^0\Big)\nn
&+&R_1^0(\la)\Big({1\over9}-{\pi^2\over18} \Big)+
   R_1^1(\la)\Big({128\over9}-{\pi^2\over72}\Big)+
   R_1^2(\la)\Big(-{100\over9}-{\pi^2\over24}\Big)\nn
&+&R_2^0(\la)\Big({7\over6}-{\pi^2\over18}\Big)+
   R_2^1(\la)\Big({16\over9}+{7\pi^2\over72}\Big)+
   R_2^2(\la)\Big({\pi^2\over24}\Big)\nn
&-&{46\over9}R_3^0(\la)-{32\over9}R_3^1(\la)-{32\over3}R_3^2(\la)\nn
&+&{40\over3}R_4^0(\la)+{40\over3}R_4^1(\la)
\co
\label{key}
\eea
where the integrals $B^{2k}$ admit a simple analytical representation
\bea
B^{2k}(\la) &=& \int_{-\infty}^{\infty}d\til y\;\til y^{2k}\,
e^{-\sqrt{1+\til y^2}\la} =
{\Gamma(k+1/2) \over \Gamma(3/2) } \left( {2 \over \la} \right)^k
K_{k+1}(\la) \co \nn
B^0(\la) &=& \,2\,K_1(\la) \co \nn
B^2(\la) &=& {2\over\la} K_2(\la)
\fs
\eea
The integrals denoted by $R_i^k$, $i=0,\ldots,4$ are defined as follows
\beq
R_i^k(\la) = N^2\,\left\{{\mr{Re}\atop\mr{Im}}\right.
\int_{-\infty}^{\infty}d\til y\;\til y^k\,e^{-\sqrt{1+\til y^2}\,\la}\,
K_i^{\pi\pi}\left(2(1+{\rm i}\til y)\right)
\qquad\mr{for}\left\{{k\;\mr{even}\atop k\;\mr{odd}}\right.\nn
\eeq
(the $K_i^{\pi\pi}$ functions appear in the $\pi\pi$ scattering
amplitude at the two--loop level \cite{BCEGS} -- there they are defined
without the $\pi\pi$ superscript -- and collected in Appendix B) and can be
easily calculated numerically, while an analytic representation is not
available.
The integrals $B^{2k}$ and $R_i^k$ are plotted in Fig.\,\ref{fig:BR}. As
can be seen there, they are all of a similar magnitude -- in the
representation (\ref{RMser}) the small parameter $\xi$ appears explicitly
such that all the remaining coefficients are not expected to show any
special hierarchy.

The last ingredient needed for an evaluation of $R_M(\Mpi,L)$ is the numerical
values of the LEC which appear in the coefficients $\bb_i$ (which are
specific for $\Nf=2$), and in $\Fpi$, if the latter needs to be computed
from $\Mpi$. The expression of the $\bb_i$ in terms of the relevant LEC
can be found in the appendix.
At order $p^2$ no LEC appears. At order $p^4$ four LEC appear:
three of them were determined rather precisely in \cite{CGL}, whereas for the
one that dictates the NLO quark mass dependence of $\mpi^2$, we rely on the
estimate in \cite{GL84}. Altogether, this means that we use
\bea
\ltil_i &\equiv & \log {\Lambda_i^2 \over \mu^2 } \nn
\Lambda_1 &=& 0.12 \stackrel{+0.04}{_{-0.03}} \; \gev \co \qquad \qquad 
\Lambda_2 = 1.20 \stackrel{+0.06}{_{-0.06}} \; \gev \co \nn
\Lambda_3 &=& 0.59 \stackrel{+1.40}{_{-0.41}} \; \gev \co \qquad \qquad 
\Lambda_4 = 1.25 \stackrel{+0.15}{_{-0.13}} \; \gev \fs
\eea

At order $p^6$ six new LEC appear. Two of them were determined in \cite{CGL},
whereas for the remaining four we rely on the resonance saturation hypothesis
\cite{BCEGS}, with a (conservative) 100\% error estimate:
\bea
\tilde r_1 &=&-1.5\times(1\pm1) \co \qquad \qquad 
\tilde r_2 = 3.2\times(1\pm1) \co \nn
\tilde r_3 &=&-4.2\times(1\pm1) \co \qquad \qquad 
\tilde r_4 =-2.5\times(1\pm1) \co \nn
\tilde r_5 &=& 3.8 \pm 1.0 \co  \qquad \qquad \qquad \; \;
\tilde r_6 = 1.0 \pm 0.1 \fs
\eea
Inserting these values and evaluating the corresponding $\bb_i$ one can
then determine the integrals $I_2,...,I_6$ in the $\xi$-expansion
(\ref{RMser}), and hence $R_M(\Mpi,L)$, the details being given in section
3 and the appendix.


\section{Numerical evaluation}


\subsection{Quark mass dependence of $F_\pi$}

In CHPT the expansion parameter is $\xi$, and in (\ref{MpiL}, \ref{RMser})
the finite volume effects have been expressed as a power series in this
parameter.  If, in a lattice calculation, both $\Mpi(L)$ and $\Fpi(L)$ have
been determined, then the square of $\Mpi(L)/(4\pi\Fpi(L))$ may be taken as
a first approximation to $\xi$, with iterative refinement through
(\ref{MpiL}, \ref{FpiL}).  In order to give a numerical prediction with
only $\Mpi$ as input, we must know how $\xi$ depends on the quark mass, or
--~equivalently~-- we are left with the problem of evaluating the pion mass
dependence of $F_\pi$.  The analytic expression of the latter is known to
next--to--next--to leading order \cite{Bijnens:1998fm}, and reads in our
variables
\bea
\Fpi&=&F\Bigg\{
1+X\Big[\Ltil+\ltil_4\Big]+X^2\Big[-{3\ovr4}\Ltil^2
+\Ltil\Big(-{7\ovr6}\ltil_1-{4\ovr3}\ltil_2+\ltil_4-{29\ovr12}\Big)
\nonumber
\\
&&\qquad\qquad\qquad\qquad\qquad\quad\;
+{1\ovr2}\ltil_3\ltil_4-{1\ovr12}\ltil_1-{1\ovr3}\ltil_2-{13\ovr192}
+\rtil_F(\mu)\Big]
\Bigg\}
\label{eq:Fpi}
\eea
where $\Ltil=\log(\mu^2/\Mpi^2)$, $X=M_\pi^2/(N F^2)$ and $\rtil_F (\mu)$
is the relevant combination of $O(p^6)$ LEC.
In order get a numerical value of $F_\pi$ at fixed $M_\pi$ we need to specify
our input parameters; the $\ltil_i$ have been given in the previous section,
whereas for the new LEC we take $\rtil_F(\mu)= 0 \pm 3$ and vary the
renormalization scale between $\mu=0.5$ and 1 GeV.
The only remaining parameter is the value of $F$, the decay constant in the
chiral limit. We fix it by inverting Eq.~(\ref{eq:Fpi}) --~now at the physical
pion mass~-- and expressing $F$ in terms of $F_\pi$.
From $F_\pi=92.4\pm0.3$ MeV we obtain
\beq
F= (86.2 \pm 0.5) \mbox{MeV} \fs
\eeq

\begin{figure}[t!]
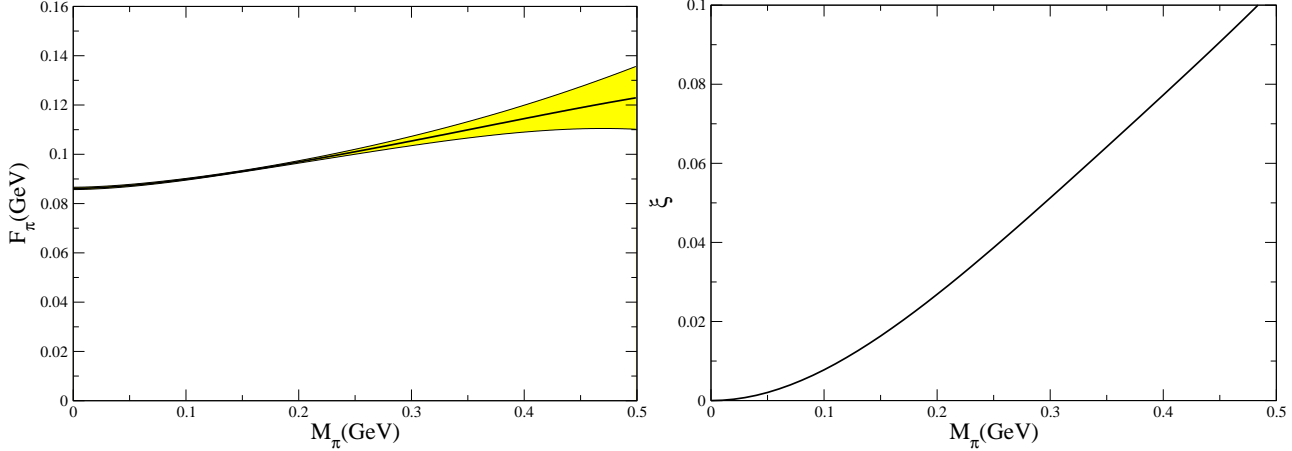

\begin{center}
\includegraphics[width=8.5cm]{Fpi_GeV.eps}
\includegraphics[width=8.5cm]{xi.eps}
\end{center}
\vspace{-6mm}
\caption{The plot on the left shows the $1\si$-band of the pion mass
  dependence of $F_\pi$. The plot on the right shows the dependence of
  $\xi$ on the pion mass, without including any uncertainty.}
\label{fig:fpi} 
\end{figure}

The pion mass dependence of $F_\pi$ is illustrated in Fig.\,\ref{fig:fpi}
and is seen to be rather mild.  This means that in the finite-volume
condition (\ref{xpt_cond_mod}) the numerical value on the r.h.s.\ (which
was assigned using the physical pion mass) holds to a good approximation
also for substantially heavier pions -- the increase of $F_\pi$ with the
pion mass is too mild to matter in this respect. In our numerical studies
of the CHPT formulae we restrict ourselves to $L\!\geq\!1.5\fm$ and
actually expect that this bound might already be too low with respect to
the condition (\ref{xpt_cond_mod}).  Let us remark that the same bound
has to be respected even if one works in the $\epsilon$--regime
(\ref{xpt_eps}) and wants to compare to CHPT formulae: in this regime
one must still have a relatively large volume and is consequently forced to
use tiny -- smaller than physical -- quark masses.

By contrast, the parameter $\xi$ remains small even for pion masses of
about half a GeV, as is shown on the right-hand side of
Fig.\,\ref{fig:fpi}.  In our numerical analysis we will use $\xi$ exactly
as given in this figure and ignore the uncertainties of this ``computed''
$\xi$ (shown only in the plot on the left-hand side of Fig.\,\ref{fig:fpi})
since, as explained above, by measuring $\Fpi(L)$ one can determine $\xi$
iteratively from the lattice data.  We do, however, determine the
uncertainty of the coefficients $I_{2m}$ in the series (\ref{RMser}), and
for that aim we use the correlation matrix among the input parameters which
can be obtained from Ref.~\cite{CGL}.  The numerical values and the details
of the analysis are given in Appendix C.


\subsection{Finite volume effects in the pion mass}

We are now in a position to evaluate the L\"uscher formula (\ref{RMser}) for
the relative mass shift $R_M(\Mpi,L)$.
Before doing so, we find it instructive to have a look at the product
\beq
F({\rm i}y)\;e^{-\sqrt{\Mpi^2+y}\;L}
\label{iofy}
\eeq
which is the integrand in the L\"uscher formula (\ref{luscher}).

\begin{figure}[t!]
\begin{center}
\begin{tabular}{rrr}
\epsfig{file=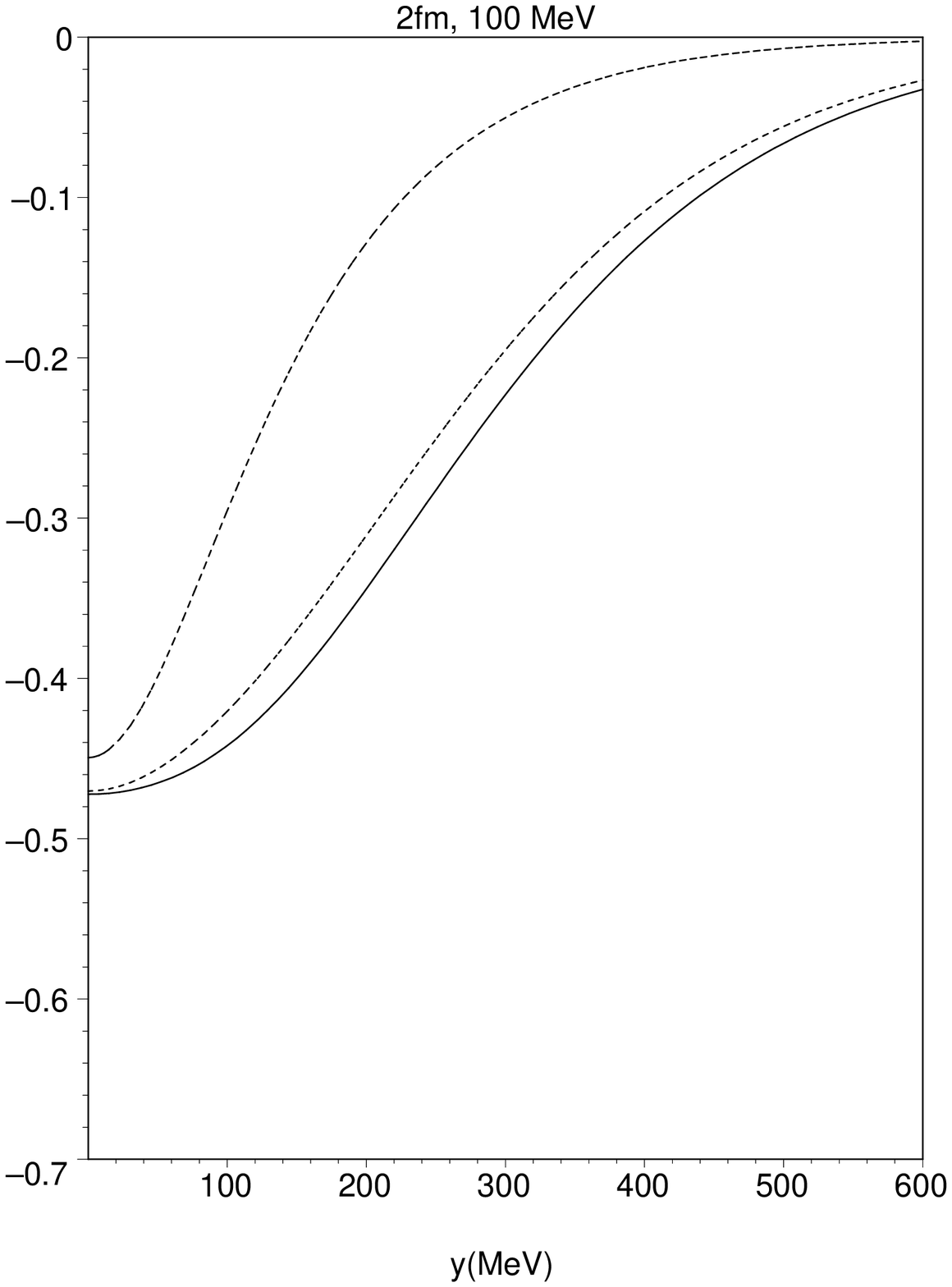,height=6.5cm}&
\epsfig{file=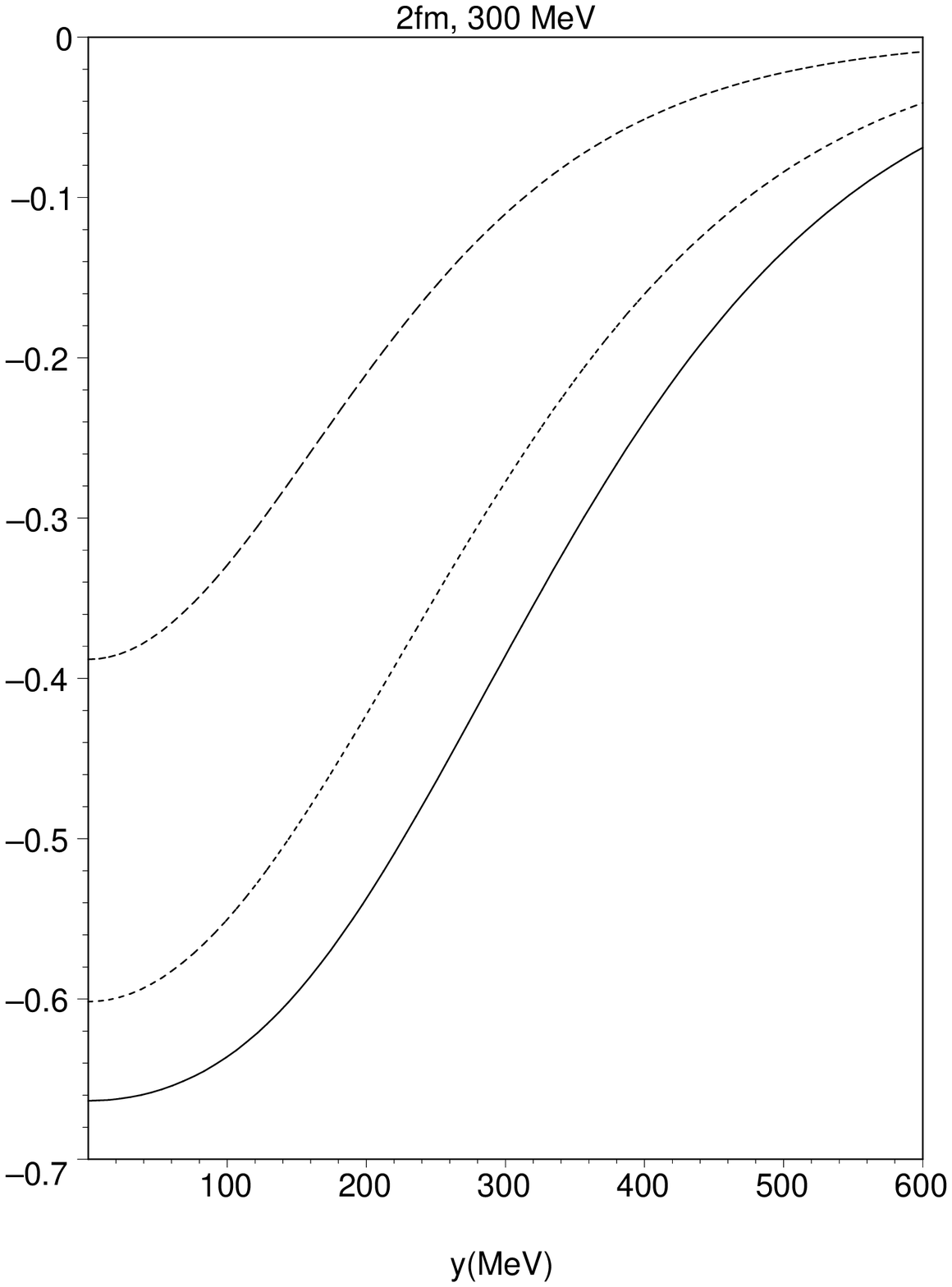,height=6.5cm}&
\epsfig{file=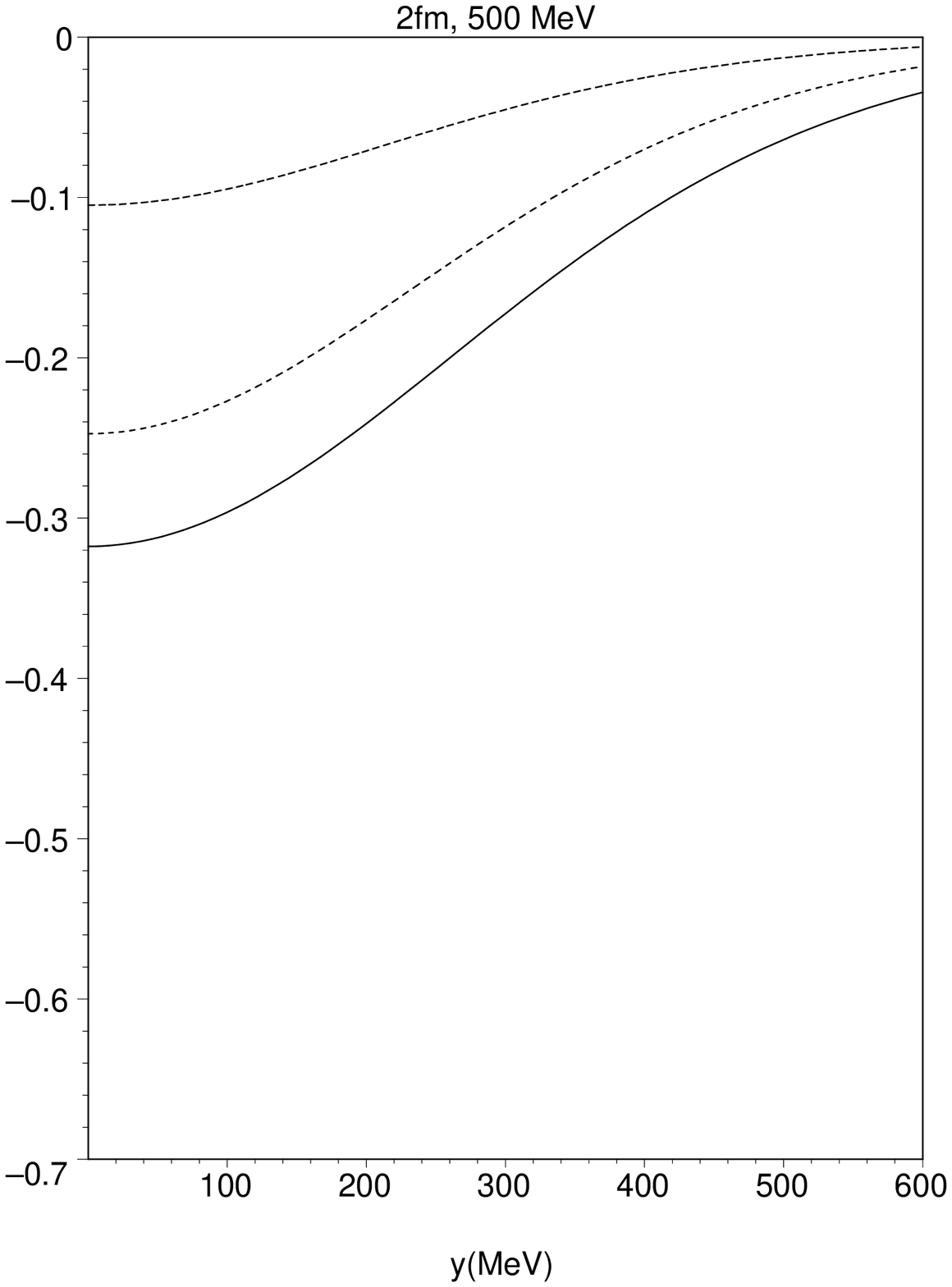,height=6.5cm}\\[4mm]
\epsfig{file=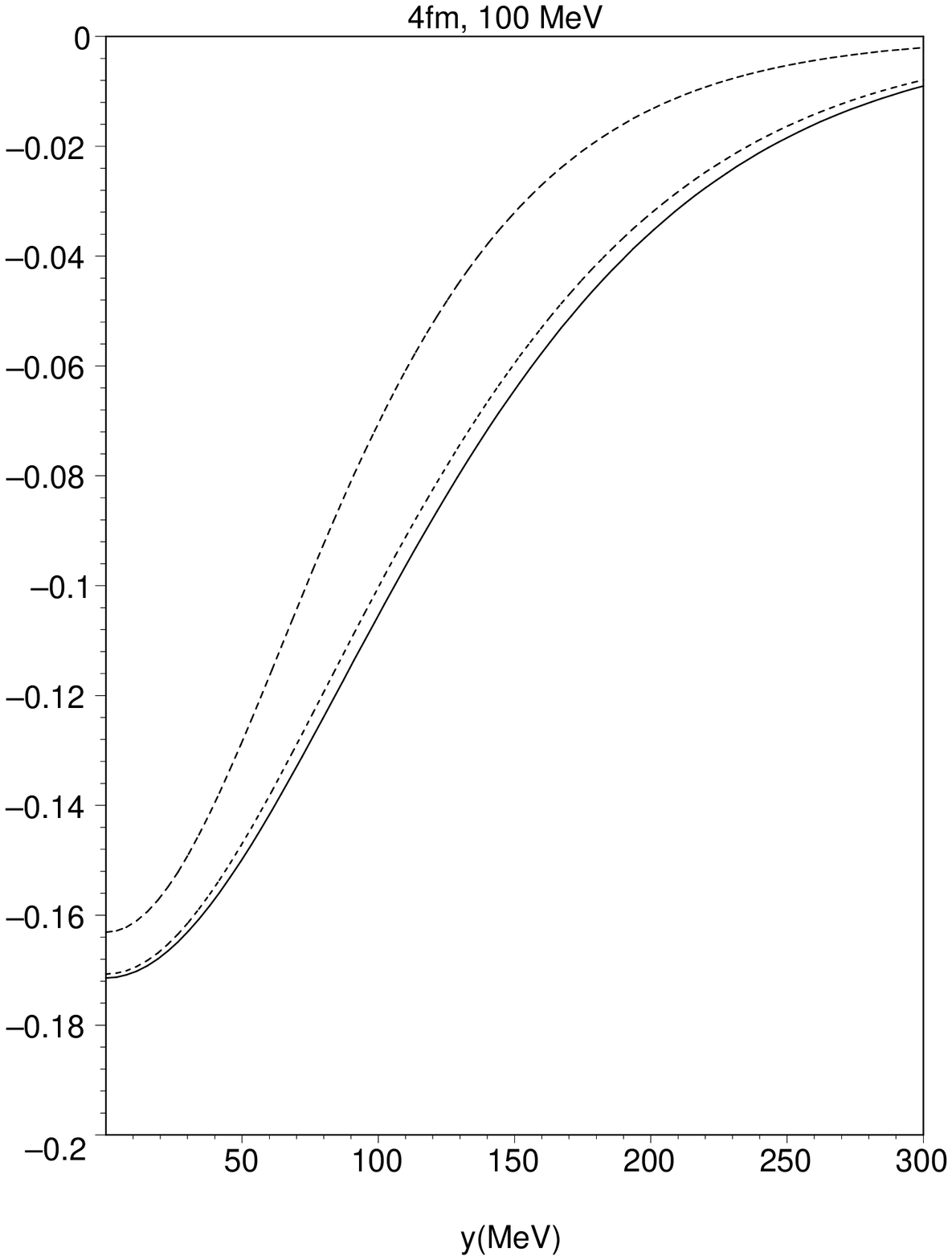,height=6.5cm}&
\epsfig{file=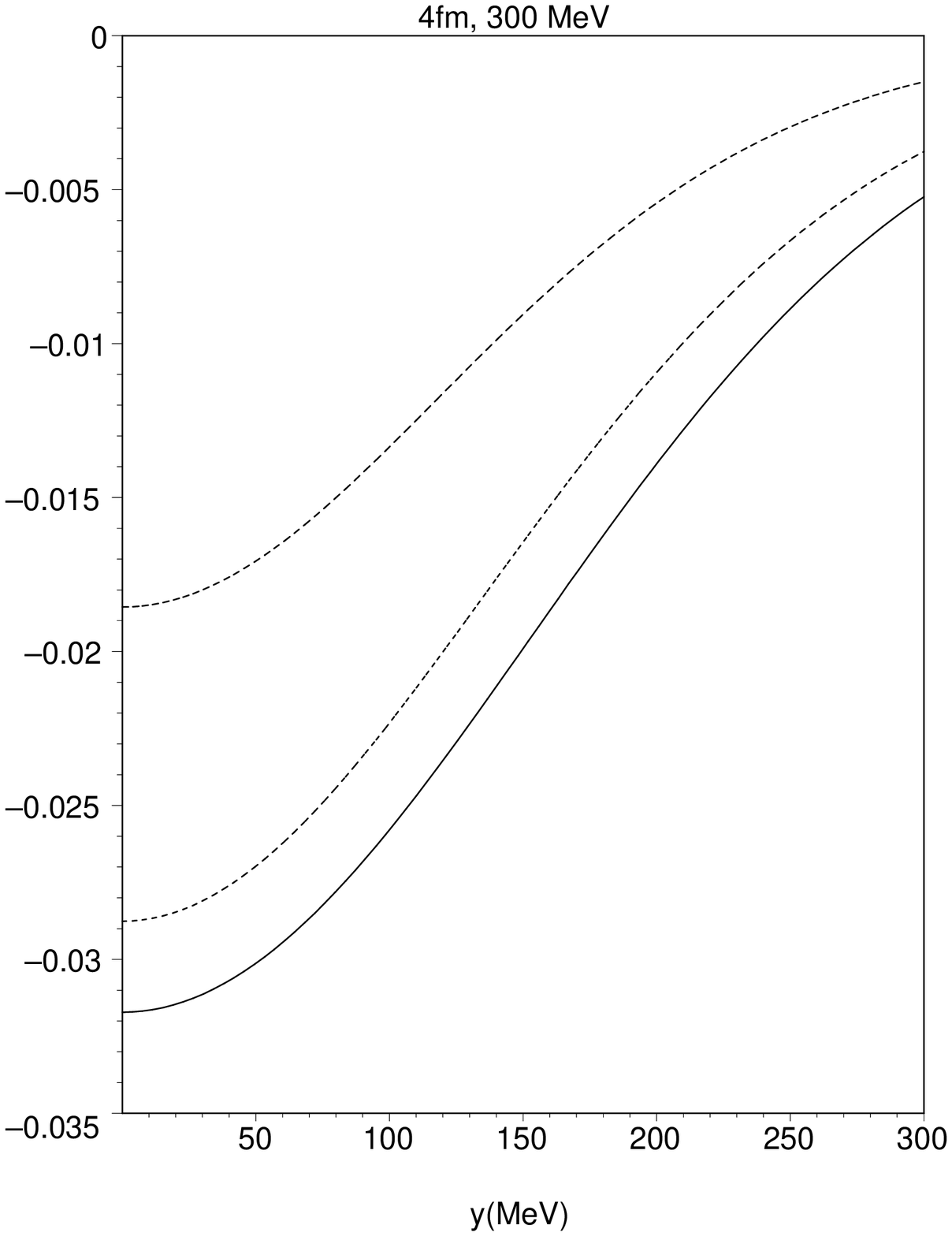,height=6.5cm}&
\epsfig{file=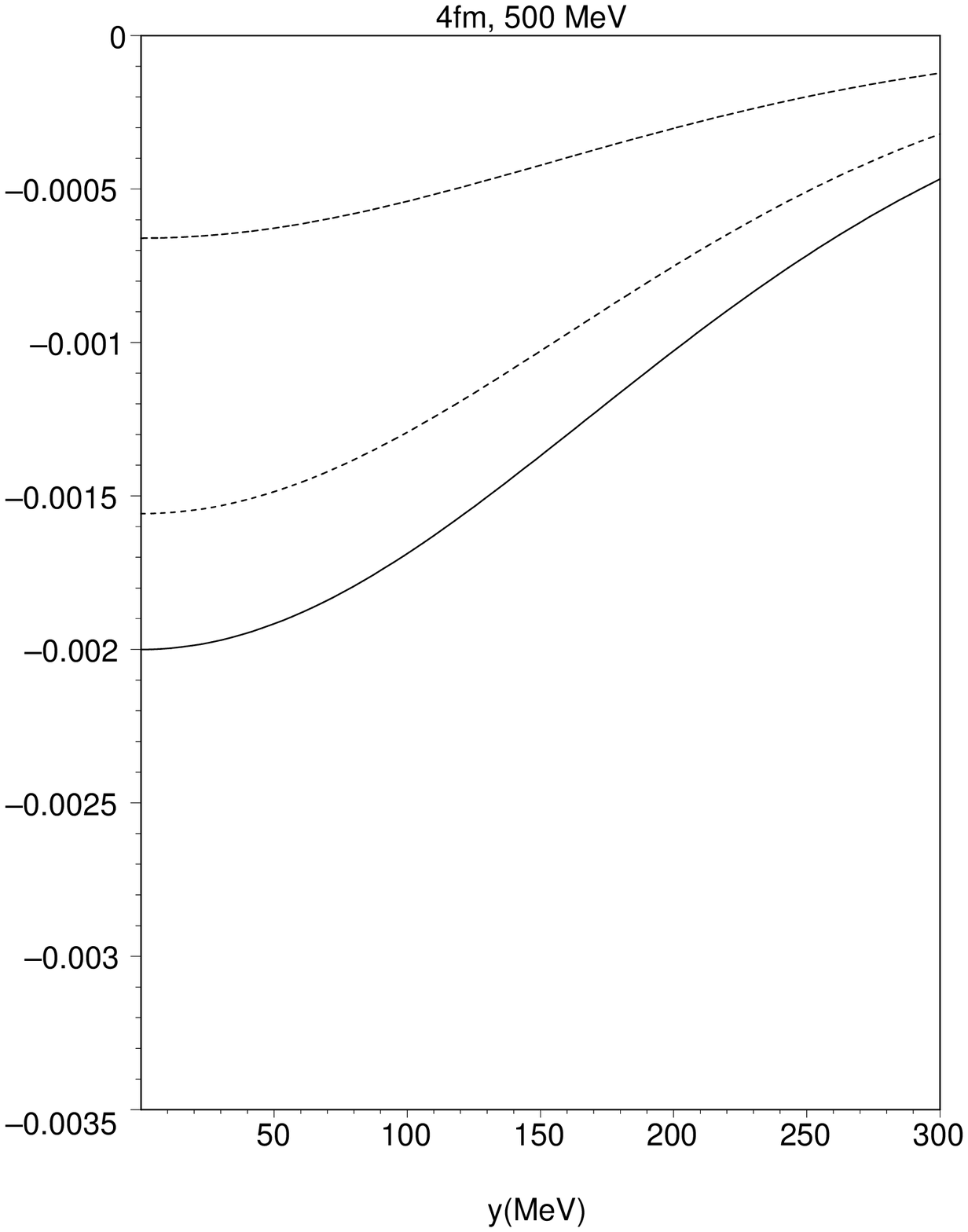,height=6.5cm}
\end{tabular}
\end{center}
\vspace{-6mm}
\caption{The integrand $I(y)=e^{-\sqrt{\Mpi^2+y^2}\,L}\,F(\mr{i}y)$ with LO,
NLO, NNLO input from CHPT. Note the change in scale in the lower panel.}
\label{fig:integrand}
\end{figure}

We have plotted this integrand in Fig.\,\ref{fig:integrand} for different
values of $\Mpi$ and $L$, evaluated at LO, NLO and NNLO in the
chiral expansion.
Since the calculation is based on CHPT and the integration variable $y$ has
the dimension of a mass, the integrand must tend to zero sufficiently fast
beyond $O(100)\MeV$ -- otherwise the outcome of the calculation cannot be
trusted. This criterion results in a clear veto against too small box sizes
(around $\sim\!1\fm$).  By the absolute amount, the two integrands are
close to each other whenever the pion is light and the box is sufficiently
large.  However, in the whole range of $\Mpi$ and $L$ shown in the
graphs, the relative difference between the LO and the NLO integrand is
substantial.  The reason is that at LO the function $F(\nu)$ is a
constant ($F_2$ can depend at most linearly on $\nu$, and since it has to
be even in $\nu$ it can only be a constant): the energy dependence of the
forward scattering amplitude --~in particular the one dictated by the
unitarity cuts~-- appears only at NLO. The difference between LO
and NLO is therefore not very useful in judging the convergence of the
chiral series, which should rather be evaluated by comparing NLO and NNLO.
The plots in Fig.~\ref{fig:integrand} show that even for $\mpi=500$ MeV the
correction in going from NLO to NNLO is not unacceptably large.
Note that in the graphs in Fig.\,\ref{fig:integrand} the product $\Mpi L$
ranges from about 1 (top left) to 10 (bottom right): in the former case the
product is too small for the asymptotic formula (\ref{luscher}) to apply --
we will confirm this below, by comparing this result to the one obtained
with one-loop CHPT in finite volume, eq.~(\ref{mpi}).

\begin{figure}[tb]
\epsfig{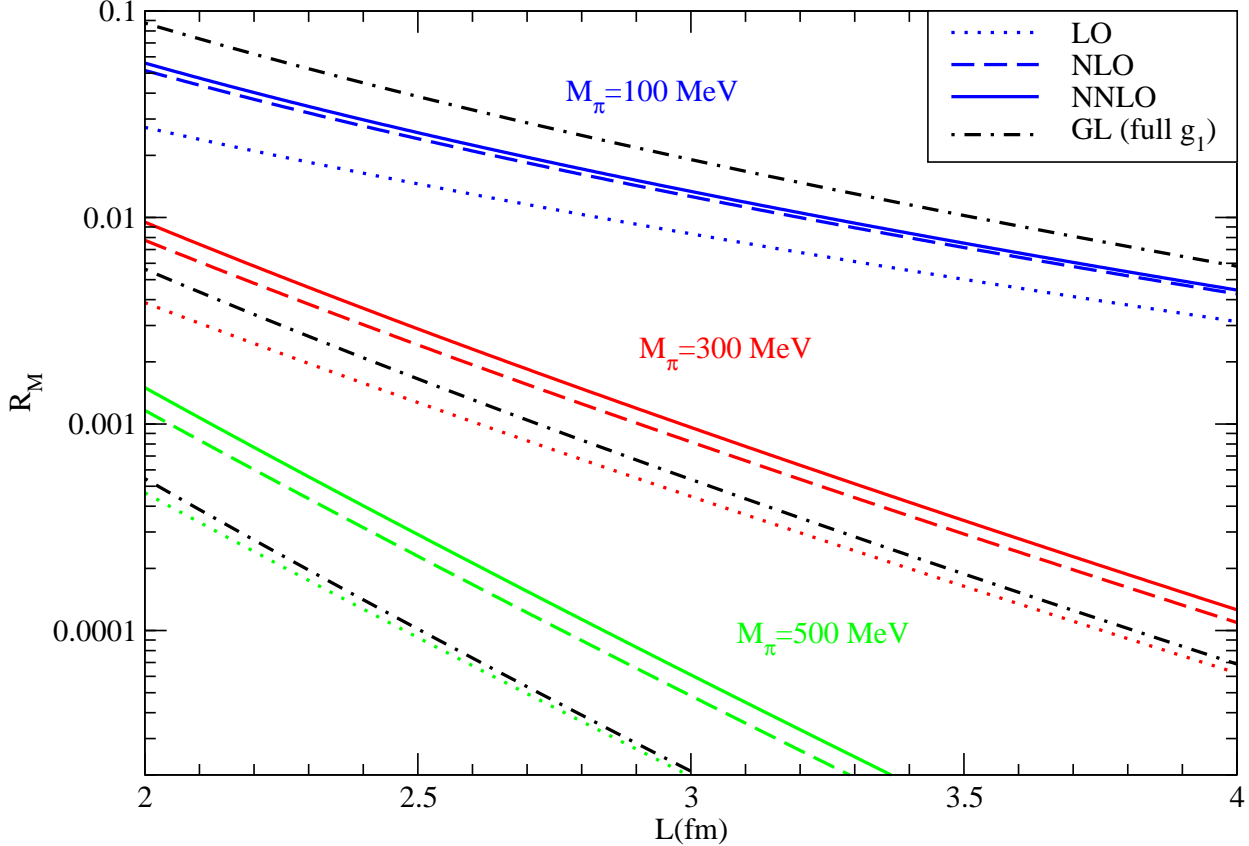}
\vspace{-5pt}
\caption{The relative mass shift $R_M(\Mpi,L)$ versus $L$ for a few values of
the (infinite volume) mass $\Mpi$. In addition, the full 1-loop result
(\ref{mpi}) by Gasser and Leutwyler is included.}
\label{fig:RvsL}
\end{figure}

\begin{figure}[tb]
\epsfig{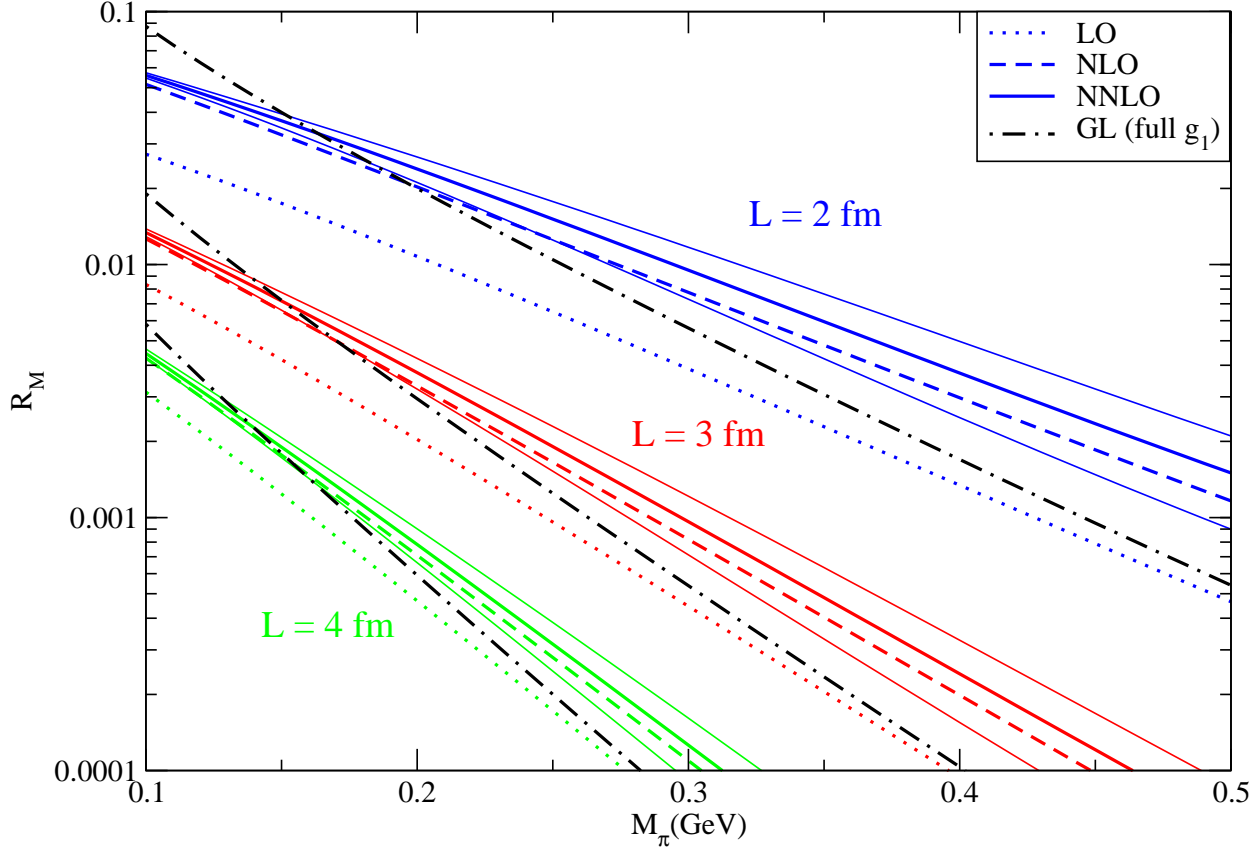}
\vspace{-5pt}
\caption{The relative mass shift $R_M(\Mpi,L)$ versus $\Mpi$ for a few values
of the box length $L$. The thin lines around the NNLO curves represent the
corresponding uncertainties.  For comparison, the full one-loop result
(\ref{mpi}) of Gasser and Leutwyler is included.}
\label{fig:RvsM}
\end{figure}

The relative mass shift $R_M(\Mpi,L)$ calculated with
the L\"uscher formula (\ref{luscher}, \ref{RMser}) is plotted in
Fig.~\ref{fig:RvsL} as a function of $L$ for different values of $\mpi$ and
in Fig.~\ref{fig:RvsM} as a function of $\Mpi$ for selected values of $L$.
In the latter figure we also show the uncertainty coming from
the LEC -- as expected the band grows with the pion mass.
These figures show that the chiral expansion converges nicely for very
light pions ($\Mpi\!=\!100$ MeV) and more slowly, but still satisfactorily for
heavier pions. For light pions, however, the comparison between the full
one-loop CHPT calculation and L\"uscher's formula (evaluated with the LO
forward scattering amplitude) shows that the leading exponential behavior
is not numerically dominating, even for volumes as large as $L\!=\!4$ fm (where
$\Mpi L\!\sim\!2$ for the lightest pion considered). Here, the
use of L\"uscher's formula is not justified, and one should rather rely on the
full one-loop result in CHPT in finite volume. On the other hand, the fact
that the NLO correction in L\"uscher's formula is rather large indicates
that even the full one-loop result does not give a reliable answer. In such
cases (e.g.\ when one will be able to simulate 200 MeV pions
in a 2 fm volume) one would need a full two-loop calculation of the pion
mass in CHPT in finite volume in order to reliably estimate the finite
size corrections.

For heavier pion masses the higher exponentials which are neglected in
L\"uscher's formula are less important and one is entitled to fully rely on
this convenient formula. We find that for masses above 200 MeV and $L \geq
2$ fm the finite size effects are at most of the order of a few percent. We
stress that in this range of masses and volumes our evaluation of the
finite size effects is reliable because we are able to check both the
convergence of the chiral expansion and that of the large volume expansion
of L\"uscher.  For $\mpi L\!\simeq\!4$ and $L\!\simeq\!4\fm$ or larger the
predicted shift is below $0.1\%$: in such cases we may conclude that for
all practical purposes the mass obtained on the lattice coincides
numerically with the infinite-volume one. Mass shifts of the order of $1\%$
are predicted only for $\Mpi\simeq200\div300\MeV$,
$L\simeq2\div2.5\fm)$. This is the region where precision tests of lattice
QCD calculations (we stress again that our calculation applies only to full
QCD) will need the application of such corrections.

\begin{table}[tb]
\centering
\begin{tabular}{|l|l|l|l|l|l|l|}
\hline
&\quad$1.5\fm$&\quad$2.0\fm$&\quad$2.5\fm$
&\quad$3.0\fm$&\quad$3.5\fm$&\quad$4.0\fm$\\
\hline
$100\MeV$
&$0.58\times10^{-1}$&$2.73\times10^{-2}$&$1.46\times10^{-2}$
&$0.83\times10^{-2}$&$5.02\times10^{-3}$&$3.13\times10^{-3}$\\
&$1.35(3)$&$5.15(14)$&$2.41(7)$&$1.26(4)$&$7.15(24)$&$4.27(15)$\\
&$1.50(2)$&$5.60(14)$&$2.58(8)$&$1.34(5)$&$7.51(29)$&$4.45(18)$\\
&$2.39$&$8.74$&$3.85$&$1.91$&$10.2$&$5.81$\\
&$3.31$&$11.6$&$4.97$&$2.41$&$12.7$&$7.13$\\
\hline
$200\MeV$
&$2.87\times10^{-2}$&$1.08\times10^{-2}$&$0.45\times10^{-2}$
&$2.04\times10^{-3}$&$0.96\times10^{-3}$&$0.47\times10^{-3}$\\
&$6.22(47)$&$2.03(18)$&$0.78(8)$&$3.31(34)$&$1.50(16)$&$0.71(8)$\\
&$7.61(73)$&$2.39(27)$&$0.90(12)$&$3.74(52)$&$1.67(24)$&$0.78(12)$\\
&$6.56$&$2.00$&$0.73$&$2.95$&$1.29$&$0.59$\\
&$11.3$&$3.31$&$1.17$&$4.65$&$2.00$&$0.91$\\
\hline
$300\MeV$
&$1.31\times10^{-2}$&$0.39\times10^{-2}$&$1.27\times10^{-3}$
&$0.45\times10^{-3}$&$1.64\times10^{-4}$&$0.62\times10^{-4}$\\
&$2.87(41)$&$0.78(12)$&$2.42(41)$&$0.82(14)$&$2.93(53)$&$1.09(20)$\\
&$3.65(77)$&$0.95(22)$&$2.89(72)$&$0.96(25)$&$3.41(92)$&$1.26(35)$\\
&$2.25$&$0.56$&$1.65$&$0.54$&$1.88$&$0.69$\\
&$4.59$&$1.12$&$3.27$&$1.05$&$3.65$&$1.32$\\
\hline
$400\MeV$
&$0.58\times10^{-2}$&$0.13\times10^{-2}$&$0.34\times10^{-3}$
&$0.94\times10^{-4}$&$0.27\times10^{-4}$&$0.79\times10^{-5}$\\
&$1.35(28)$&$0.30(6)$&$0.74(17)$&$1.98(46)$&$0.56(13)$&$1.64(39)$\\
&$1.75(55)$&$0.37(12)$&$0.91(32)$&$2.41(86)$&$0.67(25)$&$1.96(73)$\\
&$0.85$&$0.17$&$0.40$&$1.03$&$0.29$&$0.83$\\
&$2.01$&$0.41$&$0.97$&$2.50$&$0.69$&$1.99$\\
\hline
$500\MeV$
&$0.26\times10^{-2}$&$0.47\times10^{-3}$&$0.09\times10^{-3}$
&$0.20\times10^{-4}$&$0.44\times10^{-5}$&$0.10\times10^{-5}$\\
&$0.66(17)$&$1.16(31)$&$0.23(6)$&$0.48(13)$&$1.07(29)$&$0.24(7)$\\
&$0.87(34)$&$1.50(60)$&$0.29(12)$&$0.61(26)$&$1.34(57)$&$0.31(13)$\\
&$0.34$&$0.54$&$0.10$&$0.21$&$0.45$&$0.10$\\
&$0.95$&$1.58$&$0.30$&$0.62$&$1.35$&$0.31$\\
\hline
\end{tabular}
\caption{The relative finite size effect $R_M(\Mpi,L)$ for selected values of
$\Mpi$ and $L$. In each cell we give the result of the L\"uscher formula
(\ref{luscher}) with the forward amplitude at LO, NLO and NNLO accuracy
from CHPT. The fourth entry gives the GL result (\ref{mpi}, \ref{g1til2})
and the last one contains the full one-loop result shifted by the
difference between the NNLO and the LO L\"uscher formula -- this, we
believe, is a reasonable estimate of the total effect. The power of ten
given in the first entry of the cell applies to all other entries.  Note
that the first column is very likely in conflict with the condition
(\ref{xpt_cond_mod}). Note, finally, that the GL result (\ref{fpi},
\ref{g1til2}) for the relative shift of $\Fpi$ is $(-4)$ times the fourth
entry (for $\Nf\!=\!2$).}
\label{tab:res}
\end{table}

Our numerical findings are summarized in Tab.\,\ref{tab:res}, where we give
for selected values of $\Mpi$ and $L$ the relative mass shift $R_M$
computed via the L\"uscher formula (\ref{luscher}, \ref{RMser}) with LO,
NLO and NNLO input from CHPT. For comparison, we give the full one-loop
result without large $L$ expansion (\ref{MpiL}, \ref{q1MF}), due to Gasser
and Leutwyler \cite{GaLeFSE1}.  We have also combined the two results by
adding to the NNLO L\"uscher formula the series of large-$L$ suppressed
exponentials which appears in the full one-loop result but not in
L\"uscher's formula. This last figure is our best estimate of the total
finite-size correction, but a sizable difference between this and the NNLO
L\"uscher result signals the presence of large uncertainties. In such cases
(as we have argued above) a full two-loop evaluation of these corrections
would presumably settle their value.

Note that in the first column of Tab.\,\ref{tab:res} we give the finite
size effects for $L=1.5$ fm, although for such small volumes we are very
likely already in conflict with (\ref{xpt_cond_mod}), and are therefore
outside the region of applicability of CHPT.
We include this column in the hope that future high-precision lattice data
in the regime $1.5\fm\!<\!L\!<\!2\fm$ might pinpoint what
eqn.\,(\ref{xpt_cond_mod}) means quantitatively.


\section{Summary and Discussion}


Making all the necessary extrapolations for lattice QCD calculations (lattice
spacing to zero, quark masses to their physical value and volume to
infinity) at the same time would be enormously expensive in terms of
computer time and is practically unfeasible. Wherever possible one should
try to use analytical methods as an aid. As far as the finite-volume effects
are concerned, the necessary theoretical tools to control these artifacts
have been developed in the eighties by L\"uscher \cite{Luscher:1985dn} and by
Gasser and Leutwyler \cite{GaLeFSE1,Gasser:1987zq}.

Recent progress on the side of lattice calculations make now this issue
of concrete relevance, and gave us the motivation to make a thorough
numerical study of the finite size effects on the pion mass. In order to do
so we have explicitly evaluated the L\"uscher formula using as input the
forward scattering amplitude evaluated at leading, next-to-leading and
next-to-next-to-leading order in two-flavor CHPT. Pushing the chiral
expansion to such a high order has shown to be absolutely essential in
order to obtain good control on the convergence of the chiral series: for a
wide range of pion masses and lattice volumes we find that the we can
evaluate these corrections reliably. As shown first by L\"uscher, these
corrections vanish \emph{exponentially} with $\Mpi L$ and are therefore
negligible for sufficiently large masses and/or volumes. We have determined
the regions where these corrections are important for precision
calculations. Our numerical results are given in Figs.~\ref{fig:RvsL},
\ref{fig:RvsM} and in Table~\ref{tab:res} -- these are the main results
of this paper.

If $\Mpi L$ is not very large, keeping only the leading exponential in the
large volume expansion, as is the case in L\"uscher's formula
(\ref{luscher}), may not be accurate enough. In such cases one can take
into account the whole series of exponentials by working with CHPT in
finite volume. One-loop expressions for the pion mass and decay constant
are available in the literature \cite{GaLeFSE1} and, in numerical form, in
our Table \ref{tab:res}, and we have used them to estimate numerically for
which values of $\Mpi L$ is L\"uscher's formula not enough accurate. For
such situations, it is at the moment difficult to estimate the size of
finite-volume effects, and a full two-loop calculation in CHPT in finite
volume seems the only viable way to do it reliably.

\section*{Acknowledgment}

It is a pleasure to thank Rainer Sommer for his collaboration at an early
stage of this work and for a careful reading of the manuscript. This work is 
partly supported by the German DFG in SFB-TR9, by the Swiss National Science
Foundation and by RTN, BBW-Contract No. 01.0357 and EC-Contract
HPRN--CT2002--00311 (EURIDICE).


\section*{A: The coefficients $\bb_1,\,\ldots\,,\bb_6$}

For convenience we provide in this appendix the explicit
expressions for the effective coupling constants $\bb_1,\ldots,\bb_6$
which appear in the $\pi\pi$ scattering amplitude up to order $p^6$
\cite{BCEGS,CGL} in the split convention 
(\ref{bisplit}):
\bea
\bb_1^0&=&-\frac{7}{6}\Ltil
+\frac{4}{3}\ltil_1-\frac{1}{2}\ltil_3-2\,\ltil_4+\frac{13}{18}
\nn
\bb_1^1&=&-\frac{49}{6}\Ltil^2
+\Big\{-\frac{4}{9}\ltil_1-\frac{56}{9}\ltil_2-\ltil_3-\frac{26}{3}\ltil_4
-\frac{47}{108}\Big\}\Ltil+\frac{16}{3}\ltil_1\ltil_4
-\frac{1}{2}\ltil_3^{\;2}-3\,\ltil_3\ltil_4-5\,\ltil_4^{\;2}
\nn
&&+\frac{28}{27}\ltil_1+\frac{80}{27}\ltil_2-\frac{15}{4}\ltil_3
+\frac{26}{9}\ltil_4-\frac{34\pi^2}{27}+\frac{3509}{1296}+\rtil_1
\nn
\bb_2^0&=&\frac{2}{3}\Ltil-\frac{4}{3}\ltil_1+2\,\ltil_4-\frac{2}{9}
\nn
\bb_2^1&=&\frac{431}{36}\Ltil^2
+\Big\{6\,\ltil_1+\frac{124}{9}\ltil_2-\frac{5}{2}\ltil_3
+\frac{20}{3}\ltil_4+\frac{203}{54}\Big\}\Ltil-\frac{16}{3}\ltil_1\ltil_4
+\ltil_3\ltil_4+5\,\ltil_4^{\;2}
\nn
&&-4\,\ltil_1-\frac{166}{27}\ltil_2+\frac{9}{2}\ltil_3-\frac{8}{9}\ltil_4
+\frac{317\pi^2}{216}-\frac{1789}{432}+\rtil_2
\nn
\bb_3^0&=&\frac{1}{2}\Ltil+\frac{1}{3}\ltil_1+\frac{1}{6}\ltil_2-\frac{7}{12}
\nn
\bb_3^1&=&-\frac{40}{9}\Ltil^2
+\Big\{-\frac{38}{9}\ltil_1-\frac{20}{3}\ltil_2+2\,\ltil_4
+\frac{365}{216}\Big\}\Ltil+\frac{4}{3}\ltil_1\ltil_4
+\frac{2}{3}\ltil_2\ltil_4
\nn
&&+\frac{89}{27}\ltil_1+
\frac{38}{9}\ltil_2-\frac{7}{3}\ltil_4-\frac{311\pi^2}{432}
+\frac{7063}{864}+\rtil_3
\nn
\bb_4^0&=&\frac{1}{6}\Ltil+\frac{1}{6}\ltil_2-\frac{5}{36}
\nn
\bb_4^1&=&\frac{5}{6}\Ltil^2
+\Big\{\frac{1}{9}\ltil_1+\frac{8}{9}\ltil_2+\frac{2}{3}\ltil_4
-\frac{47}{216}\Big\}\Ltil+\frac{2}{3}\ltil_2\ltil_4
\nn
&&+\frac{5}{27}\ltil_1+\frac{4}{27}\ltil_2-\frac{5}{9}\ltil_4
+\frac{17\pi^2}{216}+\frac{1655}{2592}+\rtil_4
\nn
\bb_5&=&\frac{85}{72}\Ltil^2
+\Big\{\frac{7}{8}\ltil_1+\frac{107}{72}\ltil_2-\frac{625}{288}\Big\}\Ltil
-\frac{31}{36}\ltil_1-\frac{145}{108}\ltil_2+\frac{7\pi^2}{54}
-\frac{66029}{20736}+\rtil_5
\nn
\bb_6&=&\frac{5}{24}\Ltil^2
+\Big\{\frac{5}{72}\ltil_1+\frac{25}{72}\ltil_2-\frac{257}{864}\Big\}\Ltil
-\frac{7}{108}\ltil_1-\frac{35}{108}\ltil_2+\frac{\pi^2}{27}
-\frac{11375}{20736}+\rtil_6
\label{biexplicit}
\eea
where
\beq
\ltil_i\equiv\log{\Lambda_i^2\over\mu^2}\co\quad
\rtil_i=N^2 r_i^r(\mu) \co\quad
\Ltil=\log{\mu^2\over\Mpi^2}
\fs
\eeq
Note that the quark masses exclusively enter through $\xi$ and $\Ltil$; the
remaining quantities are independent thereof.


\section*{B: The functions $K_i^{\pi \pi}$}

The functions $K_i^{\pi \pi}(x)$, $i=0,\ldots, 4$ have been introduced in
\cite{BCEGS}, and we reproduce them here for convenience.
With
\bdm
z=1-\frac{4}{x} \qquad\mathrm{and}\qquad
{h}(x)=\frac{1}{N\sqrt{z}}\log\frac{\sqrt{z}-1}{\sqrt{z}+1} \;,
\edm
they read
\bea
K_0^{\pi\pi}(x)&=&{z\ovr N}h(x)+{2\ovr N^2} \qquad\Big[\;=\;{1\ovr N}\bar
J\;\Big]
\nonumber\\
K_1^{\pi\pi}(x)&=&zh^2(x)
\nonumber\\
K_2^{\pi\pi}(x)&=&z^2h^2(x)-{4\ovr N^2}
\nonumber\\
K_3^{\pi\pi}(x)&=&{Nz\ovr x} h^3(x)+{\pi^2\ovr N x}h(x)-{\pi^2\ovr 2N^2}
\nonumber\\
K_4^{\pi\pi}(x)&=&{1\ovr xz} \left[K_0^{\pi\pi}(x)+{1\ovr2}K_1^{\pi\pi}(x)+
{1\ovr3}K_3^{\pi\pi}(x)+{(\pi^2-6)x\ovr 12N^2} \right]
\nonumber\;.
\eea


\section*{C: Evaluation of the uncertainties}

In order to quantify the uncertainty of the $I_{2m}$ in (\ref{RMser}), we
need to know the correlation matrix among the LEC involved and the partial
derivatives of the $I_{2m}$ w.r.t\ the LEC.

The correlation matrix $C_{ij}$ among our ten input parameters is given in
Tab.\ \ref{tab:corr}. It has been obtained from Ref.~\cite{CGL}, which
represents so far the best determination of the LEC appearing in the $\pi
\pi$ scattering amplitude. We remark that some of our input parameters were
also used as input in \cite{CGL} and therefore are statistically
independent -- this is seen in Tab.~\ref{tab:corr} where some of the
off-diagonal matrix elements are zero.

Combining (\ref{key}) and (\ref{biexplicit}) one finds
\beq
{\pa I_4\ovr\pa\ltil_1}=4B^0-{8\ovr3}B^2 \;,\quad
{\pa I_4\ovr\pa\ltil_2}={8\ovr3}B^0-{32\ovr3}B^2 \;,\quad
{\pa I_4\ovr\pa\ltil_3}=-{5\ovr2}B^0 \;,\quad
{\pa I_4\ovr\pa\ltil_4}=-2B^0
\eeq
and
\bea
{\pa I_6\ovr\pa\ltil_1}&=&
\Big(4\Ltil+16\ltil_4+{20\ovr9}\Big)B^0
+\Big(-{40\ovr3}\Ltil-{32\ovr3}\ltil_4+{32\ovr9}\Big)B^2
-8R_0^2+{16\ovr3}R_0^3\nn
{\pa I_6\ovr\pa\ltil_2}&=&
\Big({32\ovr3}\ltil_4+{64\ovr9}\Ltil-{40\ovr27}\Big)B^0
+\Big(-{560\ovr9}\Ltil-{128\ovr3}\ltil_4+{464\ovr27}\Big)B^2
\nn
&{}&
+{80\ovr9}R_0^0-{80\ovr9}R_0^1-{128\ovr9}R_0^2-{16\ovr3}R_0^3\nn
{\pa I_6\ovr\pa\ltil_3}&=&
\Big(-15\Ltil-5\ltil_3-11\ltil_4-{3\ovr4}\Big)B^0
-5R_0^0\nn
{\pa I_6\ovr\pa\ltil_4}&=&
\Big({14\ovr3}\Ltil+16\ltil_1+{32\ovr3}\ltil_2-11\ltil_3-10\ltil_4
-{110\ovr9}\Big)B^0
\nn
&{}&
+\Big(-{160\ovr3}\Ltil-{32\ovr3}\ltil_1-{128\ovr3}\ltil_2+{448\ovr9}\Big)B^2
+{52\ovr3}R_0^0-{64\ovr3}R_0^1-{160\ovr3}R_0^2
\nn
{\pa I_6\ovr\pa\rt_1}&=&5B^0(\la)\nn
{\pa I_6\ovr\pa\rt_2}&=&4B^0(\la)\nn
{\pa I_6\ovr\pa\rt_3}&=&8B^0(\la)-8B^2(\la)\nn
{\pa I_6\ovr\pa\rt_4}&=&8B^0(\la)-56B^2(\la)\nn
{\pa I_6\ovr\pa\rt_5}&=&16B^0(\la)-48B^2(\la)\nn
{\pa I_6\ovr\pa\rt_6}&=&16B^0(\la)+16B^2(\la)
\;.
\eea

These two vectors are then used to sandwich the correlation matrix of our
input parameters.
With $\vec{x}\equiv\{\ltil_1,\ldots,\ltil_4,\rt_1,\ldots,\rt_6\}$, the
resulting uncertainty is
\beq
\Delta R_M=\xi^2\sqrt{\sum_{i,j=1}^{10}
\Big({\pa I_4\ovr\pa x_i}+\xi {\pa I_6\ovr\pa x_i}\Big)
\,C_{ij}\,
\Big({\pa I_4\ovr\pa x_j}+\xi {\pa I_6\ovr\pa x_j}\Big)}
\;,
\eeq
where $\pa I_4/\pa x_i$ is zero for $i>4$, and the rationale
for omitting the uncertainty in our evaluation of $\xi$ has been explained
in the text.

\begin{sidewaystable}
\begin{tabular}{l|rrrrrrrrrr}
{}&$\ltil_1$&$\ltil_2$&$\ltil_3$&$\ltil_4$&
$\rt_1$&$\rt_2$&$\rt_3$&$\rt_4$&$\rt_5$&$\rt_6$
\\
\hline
\\
$\ltil_1$&$3.5\,10^{-1}$&$-3.3\,10^{-2}$&$-3.0\,10^{-2}$&$6.7\,10^{-2}$&
$-8.8\,10^{-5}$&$1.7\,10^{-2}$&$-1.2$&$-2.1\,10^{-1}$&$-5.4\,10^{-1}$&
$-3.7\,10^{-2}$
\\
\\
$\ltil_2$&&$1.2\,10^{-2}$&$-3.0\,10^{-3}$&$-7.2\,10^{-3}$&$7.2\,10^{-6}$&
$3.7\,10^{-3}$&$9.8\,10^{-2}$&$-3.9\,10^{-1}$&$1.1\,10^{-2}$&$-4.6\,10^{-3}$
\\
\\
$\ltil_3$&&&$5.8$&$-5.5\,10^{-2}$&$-$&$-$&$-$&$-$&$-3.3\,10^{-2}$&
$7.0\,10^{-4}$
\\
\\
$\ltil_4$&&&&$4.8\,10^{-2}$&$-8.6\,10^{-6}$&$2.3\,10^{-3}$&$-1.2\,10^{-1}$&
$-7.4\,10^{-2}$&$-9.1\,10^{-2}$&$-2.2\,10^{-3}$
\\
\\
$\rt_1$&&&&&$2.3$&$-$&$-$&$-$&$4.8\,10^{-4}$&$-5.7\,10^{-5}$
\\
\\
$\rt_2$&&&&&&$10$&$-$&$-$&$-2.1\,10^{-2}$&$-5.8\,10^{-4}$
\\
\\
$\rt_3$&&&&&&&$18$&$-$&$2.3$&$1.6\,10^{-1}$
\\
\\
$\rt_4$&&&&&&&&$6.3$&$1.9$&$3.7\,10^{-1}$
\\
\\
$\rt_5$&&&&&&&&&$1.1$&$9.2\,10^{-2}$
\\
\\
$\rt_6$&&&&&&&&&&$1.1\,10^{-2}$
\end{tabular}
\caption{The correlation matrix of our 10 input parameters.}
\label{tab:corr}
\end{sidewaystable}


\end{document}